\documentclass[%
reprint,
amsmath,amssymb,
aps,
]{revtex4-2}
\usepackage{graphicx}
\usepackage{dcolumn}
\usepackage{bm}
\usepackage{hyperref}
\usepackage{xcolor}
\definecolor{Geng-color}{rgb}{0.11,0.66,0.11}

\begin{document}
	\title{Complete tunneling of acoustic waves between piezoelectric crystals}
	
	\author{Zhuoran Geng}
	\email{zhgeng@jyu.fi}
	\author{Ilari J. Maasilta}%
	\email{maasilta@jyu.fi}
	\affiliation{%
		Nanoscience Center, Department of Physics, University of Jyv{\"a}skyl{\"a}, P. O. Box 35, FIN-40014 Jyv{\"a}skyl{\"a}, Finland
	}%
	\date{\today}
	
	\begin{abstract}
	{\bf Abstract}\\
		When two piezoelectric solids are placed in close proximity, acoustic waves (phonons) can "tunnel" across a vacuum gap transmitting energy between the two solids. 
		Here, we demonstrate analytically that not only is such a phenomenon possible, but that a simple resonance condition exists for which complete transmission of the incoming wave is possible, physically corresponding to the excitation of leaky surface waves. 
		This result is derived for an arbitrary anisotropic crystal symmetry and orientation. We also show that the complete transmission condition can be related to the surface electric impedance and the effective surface permittivity of the piezoelectric material, making it possible to be determined experimentally. In addition, we present numerical results for the maximum power transmittance of a slow transverse wave, tunneling between identical ZnO crystals, as function of all possible crystal orientations. The results show a large range of orientations for which complete tunneling can be achieved. 
	\end{abstract}
	
	\maketitle

\section*{Introduction}
Acoustic waves (acoustic phonons) are deformations or vibrations propagating through a material medium. As such, they do not exist in vacuum, leading to the initial conclusion that it is impossible for the vacuum to transmit the energy of an acoustic wave between two separated media. 
However,  
at the atomic scale 
the 
vibrations of the nuclei can propagate via their electrical interactions through vacuum. 
Thus, a question can be raised, whether acoustic phonons can also be transmitted  across larger than atomic scale vacuum gaps through some electromagnetic mechanism. This is a relevant question, as with the advances in experimental techniques, nanometer to sub-nanometer scale vacuum gaps can be achieved\cite{Kim2015,Kloppstech2017,Cui2017,Jarzembski2022}. The possibility of such acoustic phonon "tunneling", as it is often called in the literature, has attracted a considerable amount of theoretical work in recent years to investigate possible mechanisms of the effect such as Casimir and van der Waals forces, particularly in the context of near-field heat transfer\cite{Prunnila2010,Sellan2012,Persson2011,Chiloyan2015,Budaev2011,xiong2014,Ezzahri2014,Sasihithlu2017,Pendry2016,Volokitin2019,Volokitin2020,Biehs2020,Alkurdi2020,Tokunaga2022}. 

One possible mechanism for acoustic wave tunneling is piezoelectricity, as in piezoelectric materials mechanical displacements carry along macroscopic electric fields. When an acoustic wave in a piezoelectric solid impinges on a free surface, it extends  a decaying, evanescent electric field into the vacuum \cite{Auld1973}. The length scale of this decay is determined by the wavelength of the acoustic wave, so by bringing another piezoelectric solid within a wavelength, acoustic power can be transmitted into the second piezoelectric solid across the vacuum gap. What makes this piezoelectrically mediated acoustic wave tunneling particularly attractive is its length scale: it is not fixed to be in the nanoscale, but operates on the typically much larger wavelength scale defined by the frequency (1 GHz would correspond to $\sim 5$ $\mu$m). The effect was introduced \cite{Kaliski1966,Balakirev1977} and observed \cite{Balakirev1978} long ago  (for more detailed background, see \cite{Geng2022}), 
but developed further more recently\cite{Darinskii2006,Prunnila2010,Geng2022}. In particular in Ref.\cite{Geng2022}, a general formalism was introduced that is applicable to any incident bulk wave mode for any anisotropic crystallographic orientation. One of the most interesting suggestions in Refs.\cite{Balakirev1977,Darinskii2006,Prunnila2010} is 
the possibility of unity transmission for some particular conditions, meaning that the incident wave 
could perhaps be completely transmitted into the adjacent solid. However, the discussions in Refs.\cite{Balakirev1977,Darinskii2006,Prunnila2010} 
are limited either by the simplified models used, or only show numerical results for the highest symmetry crystal orientations. Until now, no rigorous proof of complete acoustic wave tunneling has been presented, nor have generally valid complete tunneling conditions been put forward. 

In this work, we focus on the power transmittance of acoustic wave tunneling. We use the general formalism developed for piezoelectric acoustic wave tunneling in Ref.\cite{Geng2022} 
to analytically prove the existence of the complete tunneling phenomenon between two vacuum separated identical solids. In addition, a strikingly simple resonant tunneling condition is also derived, 
corresponding to the excitation of leaky surface waves. We also propose that this condition could be checked experimentally. Further discussion of the results are presented with a few numerical examples for ZnO crystals. In particular, we find our results differ from those obtained in Ref.\cite{Prunnila2010}. 

\section*{Results and Discussion}
\begin{figure}[ht]
	\centering
	\includegraphics[width=0.6\linewidth]{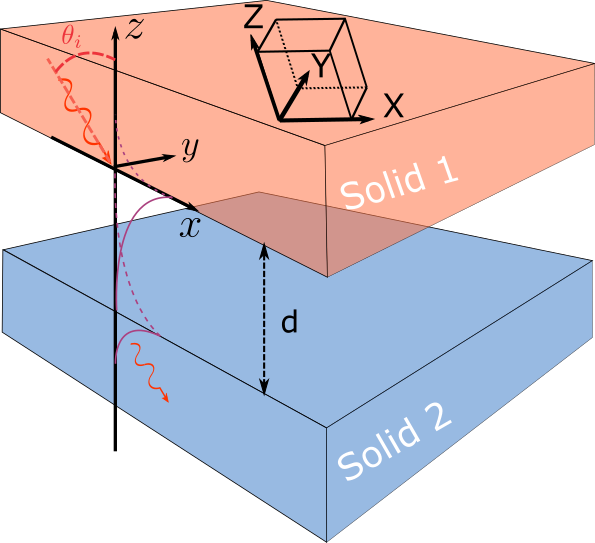}
	\caption{{\bf System under study.} Two piezoelectric solids 1, 2 are separated by a vacuum gap of width $d$. An incoming acoustic wave from solid 1 (positive $z$-axis of a laboratory coordinates $xyz$) with an incident angle $\theta_i$ tunnels across the vacuum gap into solid 2 inside the $xz$-plane.  $XYZ$ describe the intrinsic crystal coordinates, which can be rotated w.r.t. the $xyz$ coordinates.}
	\label{fig:coordinate}
\end{figure}

{\bf Tunneling of acoustic waves.}
We study a system of two anisotropic, semi-infinite piezoelectric solids separated by a vacuum gap of width $d$, as shown in Fig.\ref{fig:coordinate}. Two coordinate systems describe the relation between the crystal intrinsic orientation, denoted by $XYZ$, and the external laboratory space, denoted by $xyz$. The surfaces of the solids are assumed to be mechanically and electrically free \cite{Geng2022}, with surface normals aligned with the $z$-axis. We consider an incoming homogeneous acoustic plane (bulk) wave 
$\sim\exp(-i\pmb{k}\cdot\pmb{r}+i\omega t)$, where $\pmb{k}$ and $\omega$ are the wave vector and angular frequency, 
propagating inside the $xz$-plane (sagittal plane) from the positive $z$-axis direction towards the surface at $z=0$, 
with a positive $x$-component of wave vector ($k_x>0$). In addition, we only consider low frequency acoustic waves 
with linear dispersion and assume the usual quasistatic approximation for piezoelectric acoustic waves \cite{Auld1973} satisfying $\pmb{E}=-\nabla\Phi$, where $\pmb{E}$ and $\Phi$ are the electric field and the electric potential, respectively.

An incident bulk wave scatters into a linear combination of partial waves at an interface. These partial waves are either reflected or transmitted, and can either be homogeneous (bulk) waves or inhomogeneous (evanescent) waves bound on the surface of the solid \cite{Geng2022}. The single surface reflection and transmission coefficients, which describe the amplitudes of these scattered waves, 
can be calculated following the multiple reflection method presented in Section III.B in Ref.\cite{Geng2022}. We denote these coefficients with an overhead bar, as follows: $\bar{t}_{in\rightarrow V}^{(1)}$ is the coefficient of an incoming wave from solid 1 transmitted into a vacuum electric wave, $\bar{t}_{V\rightarrow\alpha}^{(2)}$ is the coefficient of an vacuum wave transmitted into mode $\alpha$ in solid 2, and $\bar{r}_{V}^{(i)}$ is the coefficient of an vacuum wave reflected on the vacuum side of the interface of solid $i=1,2$. In these coefficients, $\alpha=1,...,4$ correspond to the four physically allowed electroacoustic partial wave modes in the corresponding solids 1 or 2. It should be noted that these coefficients are not the direct analogs of the Fresnel coefficients\cite{born2019} from optics. 

A total transmission coefficient $t_\alpha$, which describes the amplitude ratio of a transmitted partial wave $\alpha$ in solid 2 to an incoming bulk wave from solid 1, takes a form (for derivation, see 
Section I of the Supplementary Material and Ref. \cite{Geng2022}) 
\begin{equation}
	t_\alpha=\frac{\bar{t}_{in\rightarrow V}^{(1)}\bar{t}_{V\rightarrow\alpha}^{(2)}}
	{e^{k_xd}-\bar{r}_{V}^{(1)}\bar{r}_{V}^{(2)}e^{-k_xd}}=\bar{t}_{in\rightarrow V}^{(1)}\bar{t}_{V\rightarrow\alpha}^{(2)}f_m(d)\ ,
	\label{eq:total_transmission}
\end{equation}
with $k_x$ the wave vector component along the surfaces, which is conserved in the tunneling process. This expression can be interpreted as two single surface transmission coefficients $\bar{t}_{in\rightarrow V}^{(1)}$ 
and $\bar{t}_{V\rightarrow\alpha}^{(2)}$ 
coupled by a geometrical multiple reflection factor for evanescent electrical waves in the gap $f_m(d)=[\exp(k_xd)-\bar{r}_{V}^{(1)}\bar{r}_{V}^{(2)}\exp(-k_xd)]^{-1}$ \cite{Geng2022}. It implicitly depends on the incident angle $\theta_i$ not only via $k_x=k\sin\theta_i$, but also via the coefficients $\bar{t}$ and  $\bar{r}$, which are functions of $v_x=\omega/(k\sin\theta_i)$ (
Section I in the Supplementary Material).

{\bf Derivation of the condition for complete tunneling.}
To fully describe the tunneling of the acoustic wave, we also look at the energy transfer between the solids. The time-averaged power flow density (energy flux, units [W/m$^2$]) of a transmitted partial wave in the direction normal to the surfaces (denoted as $P_\alpha$) can be obtained from the real part of the normal component of piezoelectric Poynting vector (Section I in the Supplementary Material 
). For the tunneled bulk partial waves, the transmitted power relates to the normal component of the incident power by $P_\alpha=|t_{\alpha}|^2P_{in}$, in which the input power $P_{in}$ can be from a coherent bulk wave or 
from a thermal phonon, whereas 
the reflected or transmitted evanescent partial waves in solids 1,2 are bound onto the surface and carry no power 
in the normal direction ($P_\alpha=0$ if $\alpha$ is an evanescent mode).

As there is no dissipation inside the vacuum gap, the normal direction power flow density inside the vacuum (denoted by $P_V$) is equal to the total normal direction transmitted power density (denoted by $P_{\Sigma}$). It is clear that $P_{\Sigma}$ is the sum of $P_\alpha$ over all the transmitted \textit{bulk} waves in solid 2, and we can write it using Eq.\eqref{eq:total_transmission} as $P_{\Sigma}=\sum_\alpha |\bar{t}_{in\rightarrow V}^{(1)}\bar{t}_{V\rightarrow\alpha}^{(2)}f_m(d)|^2P_{in}$, where $\alpha$ runs only over the bulk modes. The number of transmitted bulk modes can be from zero to three (in some cases four\cite{Every1992}), and if there is no bulk mode available, the power flow in both the vacuum and solid 2 are zero. On the other hand, the normal power flow inside the vacuum gap can be expressed using the Poynting's theorem under the quasistatic approximation as $P_V = 2|\bar{t}_{in\rightarrow V}^{(1)}f_m(d)|^2\mathrm{Re}[\bar{r}_V^{(2)}]P_{in}$ (see Section II in the Supplementary Material 
for the derivation). As a result, from  $P_V=P_\Sigma$ we find a relation
\begin{equation}
	2\mathrm{Re}\big[\bar{r}_V^{(2)}\big]=\sum_{\alpha=\mathrm{bulk}}|\bar{t}_{V\rightarrow\alpha}^{(2)}|^2 \ .
	\label{eq:relation_1}
\end{equation}

Furthermore, if we assume that the two solids consist of the same material with identical crystal orientations, two additional relations that link the single surface coefficients of the two solids can 
be found by exploiting the completeness of the eigensolutions of the scattering problem (see Section III in the Supplementary Material 
for the derivations). The first one relates the reflection coefficients $\bar{r}_V^{(i)}$ of the two solids as 
\begin{equation}
    \bar{r}_V\equiv\bar{r}_V^{(2)}=-\bar{r}_V^{(1)} \ .
    \label{eq:relation_2}
\end{equation}
The second one states that if the transmitted bulk wave mode $\gamma$ in the solid 2 is the same mode as the incident wave in solid 1, there exists a relation 
\begin{equation}
	\bar{t}_{\gamma\rightarrow V}^{(1)}=\bar{t}_{V\rightarrow\gamma}^{(2)}.
	\label{eq:relation_3}
\end{equation}

In addition, by comparing the relation \eqref{eq:relation_3} with Eq.\eqref{eq:relation_1}, we find the condition
\begin{equation}
	2\mathrm{Re}(\bar{r}_V)\geq|\bar{t}_{in\rightarrow V}^{(1)}|^2 \ ,
	\label{eq:relation_4}
\end{equation}
where the equality is satisfied when there exists only one transmitted bulk wave mode in solid 2 and the mode is the same as the incident wave in solid 1. By applying the relations \eqref{eq:relation_1} and \eqref{eq:relation_2}, $P_\Sigma$ can then be simplified to
\begin{equation}
	\frac{P_\Sigma}{P_{in}}=\frac{2\mathrm{Re}(\bar{r}_V)|\bar{t}_{in\rightarrow V}^{(1)}|^2}
	{4\mathrm{Re}(\bar{r}_V)^2+\big(e^{2k_xd}-|\bar{r}_V|^2\big)^2e^{-2k_xd}} \ ,
	\label{eq:total_power}
\end{equation}
which explicitly depends only on two single surface coefficients: $\bar{t}_{in\rightarrow V}^{(1)}$ and $\bar{r}_V$. 

Eq.\eqref{eq:total_power} shows that the total transmitted power $P_\Sigma$ is always less than the incident power $P_{in}$ if more than one transmitted bulk wave modes exist, since in that case the inequality Eq.\eqref{eq:relation_4} takes the "greater-than" sign. This result has the important implication that complete tunneling, i.e. the full transmission of the incident power, 
can't be achieved if the transmitted wave consists of multiple 
partial bulk waves, in contradiction to 
Ref.\cite{Prunnila2010}. 

In contrast, if there is only one transmitted homogeneous bulk mode and it is the same mode as the incident wave, then the "equal" sign of Eq.\eqref{eq:relation_4} is valid,  
and Eq.\eqref{eq:total_power} simplifies to
\begin{equation}
	\frac{P_\Sigma}{P_{in}}=
	\frac{4\mathrm{Re}(\bar{r}_V)^2}
	{4\mathrm{Re}(\bar{r}_V)^2+\big(e^{2k_xd}-|\bar{r}_V|^2\big)^2e^{-2k_xd}} ,
	\label{eq:total_power2}
\end{equation}
which very much resembles the Fabry-Perot-like form of transmission coefficients for the near-field radiative heat transfer \cite{Joulain2005}.
From Eq.\eqref{eq:total_power2}, it is clear that the maximum transmitted power is \emph{exactly} equal to the incident power ($P_\Sigma=P_{in}$) when the resonance condition
\begin{equation}
	|\bar{r}_V|=e^{k_xd} \ ,
	\label{eq:resonance_condition}
\end{equation} 
is satisfied, similar to the corresponding condition for perfect photon tunneling in near-field heat transfer\cite{Pendry1999,biehs2010a}. This proves that (i) unity transmission (complete tunneling) of an acoustic wave across a vacuum gap is possible, and (ii) the condition for it depends explicitly only on the single surface reflection coefficient $\bar{r}_{V}$, the wave vector component $k_x$ and the gap width $d$. The physical explanation of such complete tunneling 
    is the excitation of resonant coupled leaky surface waves on both interfaces (more details below and in Section V of the Supplementary material), which is fundamentally different from the principle of antireflection in optics \cite{born2019}. 

In particular, with a given material and crystal orientation, $\bar{r}_{V}$ is only a function of the incident angle and is independent of the gap width or the existence of the adjacent solid. We propose that, as a material parameter, $\bar{r}_{V}$ could be determined experimentally by measuring the effective surface permittivity $\epsilon_{\rm{eff}}(v_x)$\cite{Maugin1988,Milsom1977,Darinskii2006} or 
the TM-wave surface impedance  $Z_p(\omega,v_x)$\cite{Ingebrigtsen1969,Zhang1992} of the piezoelectric solid. They are found to be related by expressions (see Section IV in the Supplementary Material 
\begin{equation}
	\bar{r}_{V}=i\frac{\epsilon_{\rm{eff}}-\epsilon_0}{\epsilon_{\rm{eff}}+\epsilon_0},\quad 
	\bar{r}_{V}=i\frac{1+iv_x\epsilon_0Z_p}{1-iv_x\epsilon_0Z_p},
	\label{eq:vac_admittance}
\end{equation}
where $\epsilon_0$ is the vacuum permittivity. 
The effective surface permittivity concept is useful in the study of piezoelectric materials, for example for the generation and detection of acoustic waves by transducers \cite{Milsom1977} or for determining the gap wave modes between piezoelectric solids\cite{Darinskii2006}. We find the symmetric and antisymmetric gap wave conditions can be simply expressed by $\bar{r}_V=\pm i\exp(k_xd)$, which are the poles of the transmission coefficient of Eq.\eqref{eq:total_transmission} (Section IV, 
Supplemental Material). 

\begin{figure*}[t]
	\centering
	\includegraphics[width=0.9\linewidth]{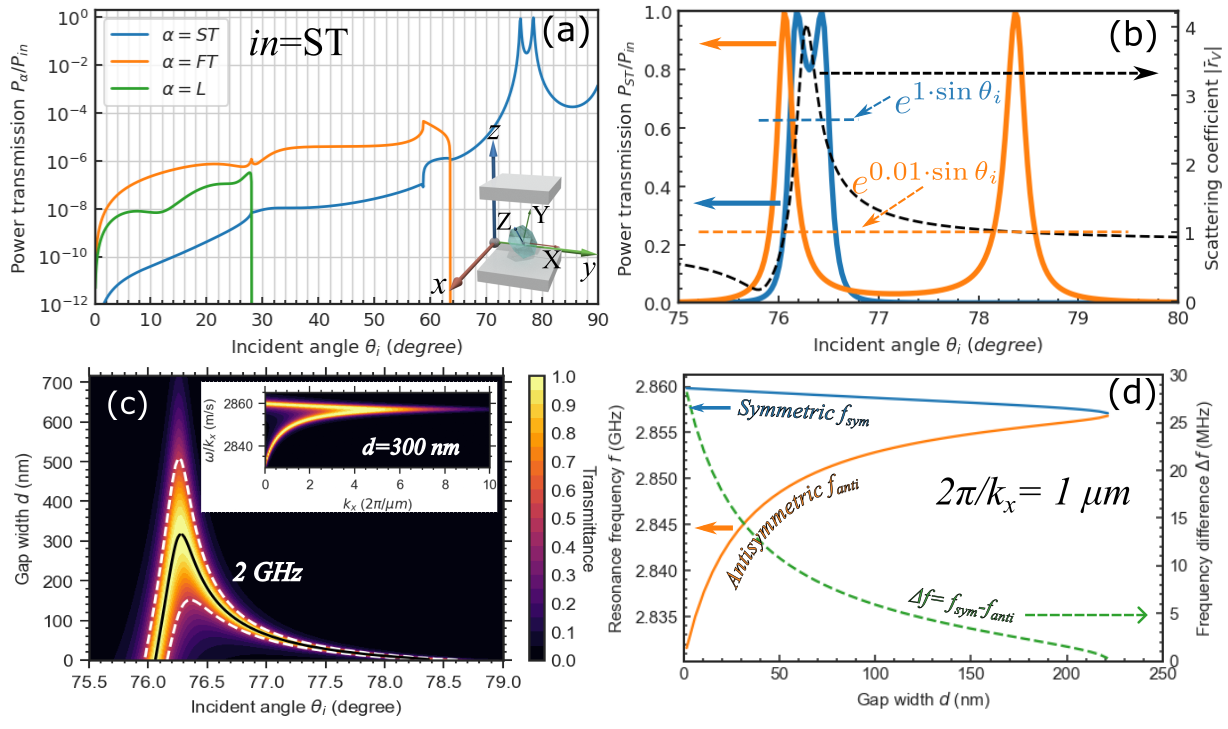}
	\caption{{\bf Angular dependence of the power transmittance of an incoming ST wave.} (a) Power transmittance $P_\alpha/P_{in}$ of the longitudinal $\alpha=L$ (green), the fast transverse $\alpha=FT$ (orange) and the slow transverse $\alpha=ST$ (blue) waves, for an incoming ST wave as function of the incident angle $\theta_i$, for two identical ZnO crystals separated by a scaled gap $kd=0.01$ and oriented with a zenith angle $\vartheta=46.89^\circ$ and an azimuth angle $\varphi=88^\circ$ (inset). We used the anisotropic crystal parameters $c_{11}=20.97\times10^{10}\ \textrm{N/m}^2$, $c_{33}=21.09\times10^{10}\ \textrm{N/m}^2$, $c_{44}=4.247\times10^{10}\ \textrm{N/m}^2$, $c_{12}=12.11\times10^{10}\ \textrm{N/m}^2$, $c_{13}=10.51\times10^{10}\ \textrm{N/m}^2$, $c_{66}=(c_{11}-c_{12})/2$, $\epsilon_{xx}=8.55\epsilon_0$, $\epsilon_{zz}=10.2\epsilon_0$, $e_{x5}=-0.48\ \textrm{C/m}^2$, $e_{z1}=-0.573\ \textrm{C/m}^2$, $e_{z3}=1.32\ \textrm{C/m}^2$, and the density $\rho=5680\ \textrm{kg/m}^3$, taken from Ref.\cite{Auld1973}. (b) Zoomed view on the peaks of the transmittance (left axis) with two values  $kd=1$ (blue solid line) and $kd=0.01$ (orange solid line). The single surface reflection coefficient $|\bar{r}_V|$ curve (black dashed line) is overlayed (right axis) together with $\exp(k_xd)=\exp(\sin\theta_i)$ (blue horizontal dashed line) and $\exp(k_xd)=\exp(0.01\sin\theta_i)$ (orange horizontal dashed line)  to demonstrate the resonance condition, Eq.\eqref{eq:resonance_condition}, for the two scaled gaps, respectively. (c) Power transmittance (color scale) as a function of  incident angle $\theta_i$ and gap width $d$ at a fixed frequency of 2 GHz (main panel), and as a function of $\omega/k_x$ and $k_x$, with a fixed $d=300$ nm (inset panel). The black solid line represents the resonance condition Eq.\eqref{eq:resonance_condition}, and the white dashed lines indicate 50\% power transmittance. (d) Frequencies of the symmetric (blue) and antisymmetric (orange) resonances as functions of gap width $d$ for a fixed $k_x$ of $2\pi/kx=1\, \mu$m, where the symmetry refers to the shape of the electrical potential function in the gap. The green dashed line shows the frequency difference between the two resonances.}
	\label{fig:ncut}
\end{figure*}

{\bf Numerical examples and physical interpretation of complete tunneling between identical ZnO crystals.}
We now turn to demonstrate the complete tunneling effect with numerical examples for two identical ZnO crystals, using the formalism developed in Ref.\cite{Geng2022}. 
The first example is shown in Fig.\ref{fig:ncut}, where the two crystals are separated with a scaled gap width of $kd=0.01$, and are both rotated first with respect to the $x$-axis by $\vartheta=46.89^\circ$ and then to the $z$-axis by $\varphi=88^\circ$ (see Ref.\cite{Geng2022} for details on the crystal rotation procedure). The mode of the incident wave in this example is chosen to be the slowest quasi-transversal wave (ST), so that there exists a critical incident angle beyond which only one bulk transmitted wave can be found, thus satisfying the general condition for complete tunneling.

In Fig.\ref{fig:ncut}(a), we plot the transmittance into each bulk mode $P_\alpha/P_{in}$ as a function of the incident angle $\theta$, where $\alpha$ can be 
the quasi-longitudinal (L), the fast quasi-transversal (FT) or the slow quasi-transversal (ST) mode, categorized based on their phase velocities. We see that for most angles, transmittance is low, except for the two sharp transmission peaks for the ST mode giving exactly unity transmission at angles between $75^\circ$ and $80^\circ$.  Abrupt cut-offs are visible for the transmitted L and FT modes, corresponding to the critical incident angles  $\theta_{L_c}\approx28^\circ$ and $\theta_{FT_c}\approx63.5^\circ$. Beyond these critical angles, the corresponding modes become evanescent, bound on the surface of the solid with no direct energy transmission into the bulk.

Fig.\ref{fig:ncut}(b) provides a zoomed view on the resonant transmission peaks, now with two different scaled gap values $kd=1$ (blue solid line) and $kd=0.01$ (orange solid line), with an overlay of the $|\bar{r}_V|$ curve 
(black dashed line), helping us also to understand the doublet structure. The two additional horizontal dashed lines represent the values of the RHS of Eq.\eqref{eq:resonance_condition} for the two $kd$ values, whereas the dashed black curve represents the LHS of Eq.\eqref{eq:resonance_condition}. 
It is clear that the unity transmission occurs where the resonance condition is valid, proving consistency between the analytical theory and the numerical approach. 
In addition, we see that with the increase of the scaled gap width from $0.01$ to $1$, the separation of the peaks is reduced, and with a further increase the two solutions would merge into one at the maximum of $|\bar{r}_V|$. With this particular ZnO crystal orientation, this maximum is about 4 as shown in the plot, which leads to a maximum gap width of $kd\approx1.4$ to observe complete tunneling (merged unity transmission peak). 
For ZnO (ST wave velocity $v=2780$ m/s), and a $2\,\mathrm{GHz}$ frequency relevant for device applications, 
this corresponds to a quite long physical distance of $d=300\,\mathrm{nm}$ with the parameters and the orientation used in the example.   

In addition, it is also useful to briefly discuss how sensitive the power transmittance is to deviations from the resonance condition, based on the above numerical example. 
    For a fixed frequency of 
    2 GHz, the transmittance as a function of both the incident angle $\theta_i$ and the gap width $d$ is presented in Fig.2(c). 
    We see that when the vacuum gap is small, e.g. $d<100$ nm, the two resonances 
    are well separated and 
    are sensitive to deviations of both $d$ and $\theta_i$. For example, 
    a 50\% drop in transmittance (white dashed lines) occurs within a 10 nm change in $d$ 
    or a 0.1 degree change in 
    $\theta_i$ for the right 
    branch. However, the merging of the two resonances leads to a higher deviation tolerance. $\theta_i=76.4^\circ$ is an interesting example, where the transmittance remains higher than 50\% for a wide range 150 nm $< d <$ 500 nm, significantly relaxing the constraint for the gap width control in measuring the tunneling. Similar discussion can also be applied 
    for a fixed gap width, for which the transmittance becomes a function of the angular frequency $\omega$ and the in-plane wave vector $k_x$, as illustrated in the inset of panel (c).

Furthermore, it is possible to take advantage of the resonances in experiments and potential applications, such as the precise control of a gap distance. By exciting a bulk wave with a known in-plane wave vector $k_x$ (for example with an interdigital transducer of finger spacing $2\pi/k_x$ \cite{Milsom1977}),    
    the frequencies of the two tunneling resonances 
    become functions of the gap distance, as demonstrated for our numerical example case in panel (d) of Fig.2. 
    We see that while the higher-frequency resonance $f_\textrm{sym}$ depends weakly on $d$, 
    the lower-frequency resonance $f_\textrm{anti}$ 
    is highly sensitive to $d$, 
    with 
    $\partial f/\partial d\approx0.3$ MHz/nm at $d<20$ nm. In addition, the frequency difference $\Delta f=f_\textrm{sym}-f_\textrm{anti}$ (green dashed line, right axis scale) 
    is also sensitive to 
    $d$, reaching a responsivity $\sim0.5$ MHz/nm at $d<20$ nm. Such relations can be envisioned to be used not only to experimentally demonstrate acoustic wave tunneling, but also to control a buried nanoscale gap distance with nanometer accuracy.  

\begin{figure}[t]
	\centering
	\includegraphics[width=1\linewidth]{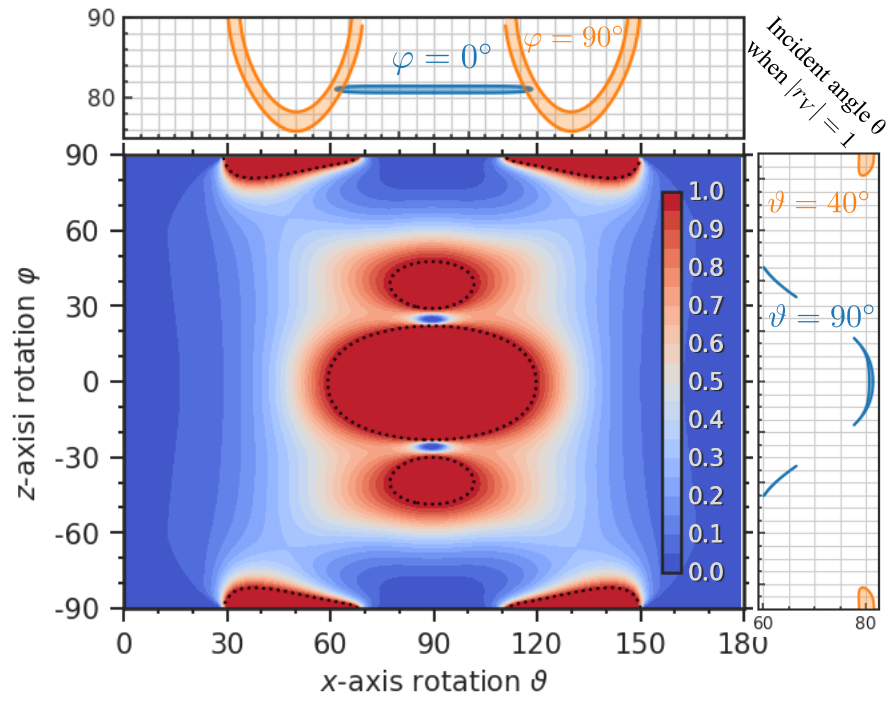}
	\caption{{\bf Crystal rotation map for complete tunneling between ZnO crystals.} Color scale of the ST-to-ST mode maximum power transmittance $P_{ST}/P_{in}$ over all incident angles, plotted as function of crystal rotation angles $\vartheta$ and $\varphi$ for anisotropic ZnO. The dotted lines encircle the regions where $|\bar{r}_V| > 1$. The panels on the top and on the right show the range of $\theta_i$ where $|\bar{r}_V| > 1$ for two fixed $\varphi$ (top), or $\vartheta$ (right). 
	Here, we fix $k_xd=kd\sin\theta_i$ instead of $kd$, as complete tunneling can be achieved by tuning $k_x$ either by 
changing the incident angle $\theta_i$ or by the angular frequency $\omega$.
	}
	\label{fig:fullrotation}
\end{figure}

In general, the complete resonant tunneling can take place for a range of crystal orientations. 
In Fig.\ref{fig:fullrotation} we show the numerically calculated  maximal power transmittance $P_{ST}/P_{in}$ (over all $\theta_i$) of an incident ST mode to a transmitted ST mode, as a function of all possible crystal rotations \footnote{ZnO has a crystallographic $6mm$ system with uniaxial symmetry\cite{Auld1973}, hence all the unique orientations of the crystal can be represented by the direction of the crystal $c$-axis ($Z$-axis) using a zenith angle $\vartheta\in(0^\circ,180^\circ)$ and an azimuth angle $\varphi\in(-90^\circ,90^\circ)$.}, using again parameters for anisotropic ZnO\cite{Auld1973} and a fixed scaled gap $k_xd=0.01$. We find a significant parameter space  for orientations, with multiple separate regions, where complete tunneling is possible  (dark red regions). 
To validate the consistency of the numerics with the analytical condition, Eq.\eqref{eq:resonance_condition}, we also plot a set of dotted contour lines in Fig.\ref{fig:fullrotation} to encircle the orientations satisfying $|\bar{r}_V|>1$, where unity transmission is possible, finding excellent agreement. 
Another observation is that the incident angle $\theta_i$ satisfying complete tunneling varies for different crystal orientations, reaching as low values as 60$^\circ$ in some cases (right panel). 

To understand the physics, we first consider 
the three ellipsoidal unity transmission areas around $\vartheta=90^\circ$ [Fig.\ref{fig:fullrotation}]. 
Inside these areas, 
the incident ST waves are not pure shear waves and therefore couple to the other partial waves (L, FT) at the surface. As a result, when the incoming ST wave has an incident angle beyond the critical angle of the FT mode, the reflected FT wave becomes evanescent, 
with its energy concentrated on the surface. For those orientations the FT-mode waves are predominately polarized in the direction of the $c$-axis, the direction of the piezoelectric dipole, creating 
a strong piezoelectric response. 
That excites large electric potential differences on the surface and hence gives rise to a strong electric coupling across the gap, which finally enables the resonant transmission.  
On the other hand, when the azimuth rotations approach $\varphi=\pm90^\circ$ with $\vartheta=90^\circ$, the $c$-axis aligns with the $x$-axis and the ST mode becomes a pure shear mode, polarized perpendicular to the sagittal plane. Then the incident ST waves are very weakly piezoelectric, 
and also decouple from all other partial modes.

Other interesting features can also be observed in Fig.\ref{fig:fullrotation}. Nodes having low transmission at around $\varphi=\pm25^\circ$ and $\vartheta=90^\circ$ appear. This is because the electric potential excited by the reflected FT wave mode change polarity around these nodes, leading to minimized potential differences and weak coupling between the two surfaces. In addition, unity transmission is also observed in four small areas around $\varphi=\pm90^\circ$, where the single surface reflection coefficient $\bar{r}_{ST\rightarrow FT}^{(1)}$ of the reflected FT partial waves 
increases significantly (not shown). This indicates an enhanced mode conversion between the ST and FT partial wave modes at these orientations,  providing  large electric potential differences 
on the solid-vacuum interface
again via the evanescent FT wave, leading to strong tunneling signal.
A more detailed discussion of the physical interpretation of the resonance can be found in Section V of 
the Supplemental Material.

Our numerical formalism can also be applied to the particular case studied with a simplified model in Ref.\cite{Prunnila2010}, the details of which can be found in Section VI of the Supplementary Material. 
We do not find complete tunneling for the incoming modes and the crystal orientation in question, in contradiction to  Ref.\cite{Prunnila2010}.

\section*{Conclusions}

In conclusion, we have analytically and numerically proven it is possible for acoustic waves to completely tunnel across a vacuum gap between two piezoelectric solids, up to gap sizes of about a wavelength. We showed that such complete tunneling, with unity power transmittance, is possible only if one transmitted partial bulk mode is excited, it being the same mode as the incident wave. We derived a strikingly simple resonance tunneling condition for the complete tunneling effect, Eq.\eqref{eq:resonance_condition}, and proved its validity  and range of applicability with numerical examples for arbitrarily rotated ZnO crystals. As this is a strong and not a rare effect, it could have an impact in future acoustic wave devices, as well as in other application areas concerning phonons, such as controlling heat transport, optomechanics and quantum information science. 


{\bf Data availability}
All relevant data are available from the authors upon request.

{\bf Code availability}
All relevant code for simulations are available from the authors upon request.


%

{\bf Acknowledgements}\\
	This study was supported by the Academy of Finland project number 341823 and by the European Union’s Horizon 2020 research and innovation program under the grant agreement number 800923 (SUPERTED).\\ 

{\bf Author contributions}\\
Z. G. and I. M. conceived the idea, carried out the analytical derivations and wrote the manuscript. Z.G. carried out all the numerical calculations.\\

{\bf Competing interests}\\
The authors declare no competing interests.
	
\end{document}